\documentclass[
	a4paper, 
	10pt, 
	unnumberedsections, 
	two side, 
]{LTJournalArticle}

\runninghead{} 

\footertext{\textit{RISC-V Summit Europe, Barcelona, 5-9th June 2023}} 

\title{Vitamin-V: Virtual Environment and Tool-boxing for Trustworthy Development of RISC-V based Cloud Services
} 

\author{%
	A. Arelakis\textsuperscript{9}, J.M. Arnau\textsuperscript{7}, J. L. Berral\textsuperscript{5,10}, A. Call\textsuperscript{5},R. Canal\textsuperscript{10}\thanks{Corresponding author: \href{mailto:ramon.canal@upc.edu}{\tt ramon.canal@upc.edu}\\This work was supported by the Vitamin-V project (Project number: 101093062) funded by the European Union. Views and opinions expressed are, however, those of the author(s) only and do not necessarily reflect those of the European Union or the HaDEA. Neither the European Union nor the granting authority can be held responsible for them.} , S. Di Carlo\textsuperscript{1}, \\ J. Costa\textsuperscript{10},   D. Gizopoulos\textsuperscript{2}, V. Karakostas\textsuperscript{2}, F. Lubrano\textsuperscript{4}, K. Nikas\textsuperscript{6}, Y. Nikolakopoulos\textsuperscript{9}, \\B. Otero \textsuperscript{10}, G. Papadimitriou\textsuperscript{2}, I. Papaefstathiou\textsuperscript{3}, D. Pnevmatikatos\textsuperscript{6}, D. Raho \textsuperscript{8}, \\A. Rigo \textsuperscript{8}, E. Rodríguez\textsuperscript{10},  A. Savino\textsuperscript{1}, A. Scionti\textsuperscript{4}, N. Tampouratzis\textsuperscript{3}, A. Torregrosa\textsuperscript{7}}

\date{\footnotesize
\textsuperscript{\textbf{1}}Politecnico di Torino, Italy; 
\textsuperscript{\textbf{2}}University of Athens, Greece;  
\textsuperscript{\textbf{3}}Exapsys, Greece; 
\textsuperscript{\textbf{4}}LINKS Foundation, Italy;
\textsuperscript{\textbf{5}}Barcelona Supercomputing Center, Spain;
\textsuperscript{\textbf{6}}Institute of Communication \& Computer Systems, Greece;
\textsuperscript{\textbf{7}}Semidynamics, Spain;
\textsuperscript{\textbf{8}}Virtual Open Systems, France;
\textsuperscript{\textbf{9}}ZeroPoint Technologies AB, Sweden;
\textsuperscript{\textbf{10}}Universitat Politècnica de Catalunya, Spain}

\renewcommand{\maketitlehookd}{%
\begin{abstract}
\noindent Vitamin-V is a 2023-2025 Horizon Europe project that aims to develop a complete RISC-V open-source software stack for cloud services with comparable performance to the cloud-dominant x86 counterpart and a powerful virtual execution environment for software development, validation, verification, and test that considers the relevant RISC-V ISA extensions for cloud deployment. 	\end{abstract}
}

\begin{document}

\maketitle 


\section{Introduction}

RISC-V is an open-source instruction set architecture (ISA) designed to be simple, modular, and extensible \cite{Waterman:EECS-2016-129}. 
RISC-V processors offer advantages over proprietary processor architectures, including open standards, customizable designs, and lower licensing costs \cite{greengard2020will}. Yet, some challenges must be addressed before RISC-V can be widely adopted for cloud applications. 

\textbf{Ecosystem maturity:} while RISC-V has gained significant momentum in recent years, the ecosystem around RISC-V processors is still developing. This includes hardware and software tools and the number of vendors and support services available.

\textbf{Performance:} while RISC-V processors can offer good performance, they may not yet be able to match the performance of more established architectures like x86 or ARM in specific applications. This can be a barrier to adoption in performance-sensitive cloud applications.

\textbf{Compatibility:} many cloud applications are designed to run on x86 or ARM architectures and might not be compatible with RISC-V processors. This can limit the use of RISC-V in specific cloud environments.

\textbf{Security:}	as RISC-V processors become more widely adopted, they may become a target for security attacks. Ensuring the security of RISC-V-based cloud applications will be an important challenge.

\textbf{Standardization:} while RISC-V is an open standard, there is still a need for further standardization in areas such as memory management and I/O interfaces. This can lead to compatibility issues between different RISC-V implementations and limit the portability of RISC-V-based cloud applications.

Vitamin-V operates in this context and tries to address these challenges by deploying a complete RISC-V hardware-software stack for cloud services based on cutting-edge cloud open-source technologies for RISC-V cores, focusing on EPI cores.
Vitamin-V incorporates an innovative RISC-V virtual execution environment providing hardware emulation, simulation, and FPGA prototyping to enable software development, verification, and validation before actual hardware is released. Vitamin-V also contributes to porting the complete cross-compiling toolchain, software stack, and essential application libraries for the forthcoming release of the RISC-V EPI processors \cite{EPI,10.1145/3310273.3323432}.


\section{Concept and Methodologies}

\begin{figure}[hbt]
  \includegraphics[width=7.5cm]{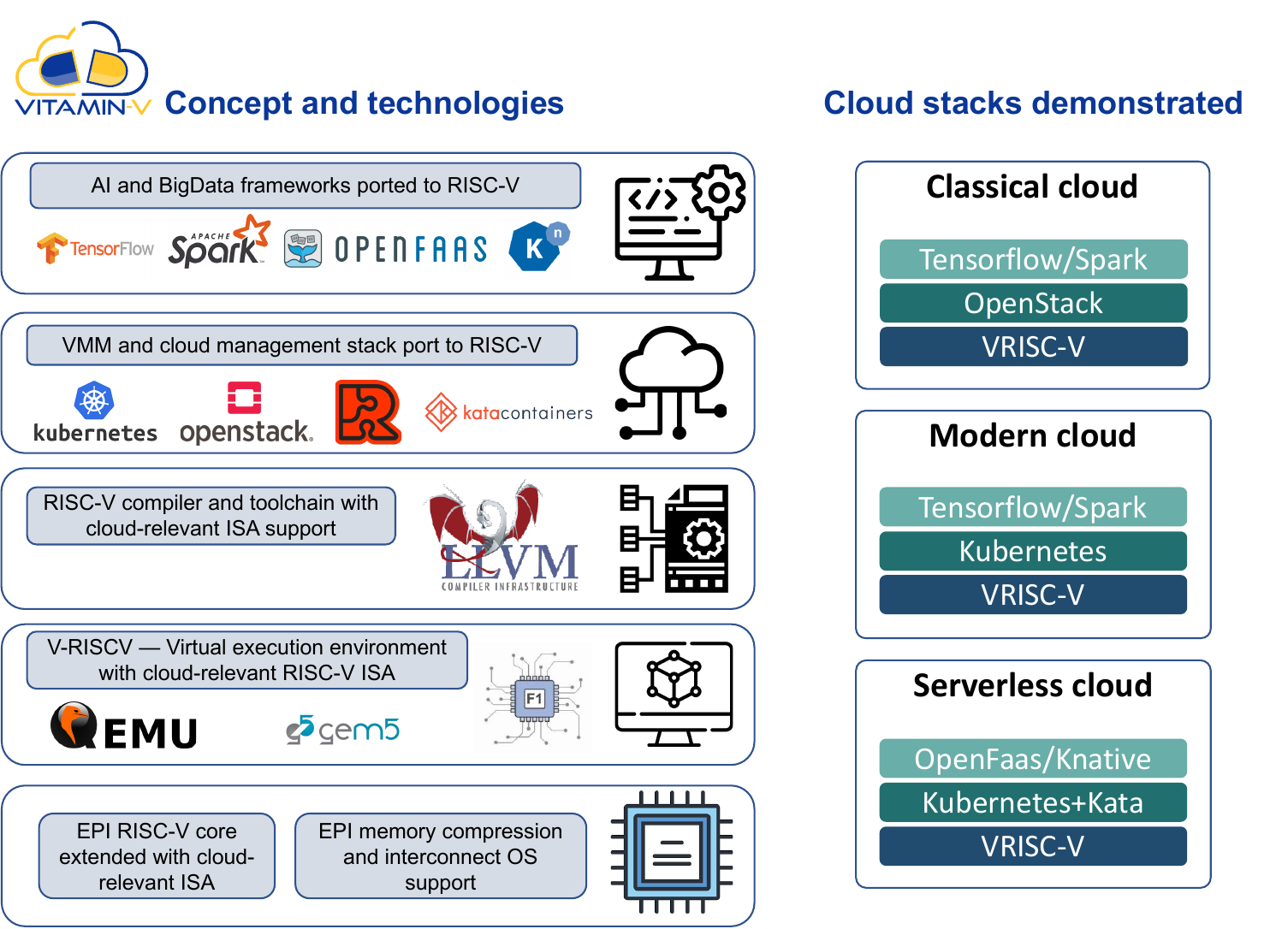}
  \caption{VITAMIN-V concept and architecture}
  \label{fig:concept}
\end{figure}

As depicted in Figure \ref{fig:concept}, Vitamin-V aims to develop a complete RISC-V cloud software stack with comparable performance to the cloud-dominant x86 counterpart.

The unavailability of full-fledged RISC-V systems poses a significant challenge for porting and evaluating cutting-edge cloud setups and software stacks. Commercial cloud systems use hardware features partially available in RISC-V virtual environments and commercial hardware cores. These features include virtualization, cryptography, and vector extensions. One of the primary goals of Vitamin-V is to overcome this challenge by building the \emph{Vitamin RISC-V virtual execution environment} (VRISC-V). VRISC-V is a complete advanced multi-layered high-performance RISC-V virtual environment based on three cutting-edge technologies: functional emulation (QEMU \cite{QEMU}), cycle-accurate simulation (gem5) \cite{gem5}, and an FPGA-based hardware prototype system node capable of running using AWS EC2 F1 FPGAs   (and thus scalable to 100s of nodes)(FPGA \cite{Amazon}). Every technology provides unique features that are important during software development, validation, verification, and testing, to favor RISC-V adoption by software developers. With QEMU and the RISC-V on FPGA, the project aims at simulating multi-node systems with more than 100 cores. The goal is to simulate multi-node systems in gem5 with cycle-accurate accuracy and microarchitectural configuration flexibility. 

Besides the RISC-V main CPUs, EPI also develops accelerators that are especially interesting for cloud setups, including one for memory compression. Consequently, Vitamin-V will support such accelerators in the VRISC-V virtual execution environment. 

Vitamin-V will fully support rapid software development in new RISC-V hardware setups. It will provide a mature compiler toolchain based on LLVM to handle the complete RISC-V ISA, including its extensions. Moreover, the project will develop a validation, verification, and testing (VVT) toolset for developers to identify software bugs and illegal or malicious code sequences, thus favoring software layers' trustworthiness in future RISC-V systems. VVT tools will use the VRISC-V platform to foster software porting and prototyping before mature hardware is released and available. This is crucial to favor the migration toward new RISC-V servers in the cloud.

Vitamin-V will port all necessary machine-dependent modules in relevant open-source cloud software distributions to enable the execution of complete cloud stacks on the VRISC-V virtual execution environment. Targeted modules include support for running entire Virtual Machines (VMs), containers, and lightweight VMs (KVM, QEMU, Docker, RustVMM), safety-security trusted execution environments (VOSySMonitoRV), cloud management software (OpenStack, Kubernetes, Kata Containers), and AI and Big Data libraries (Tensorflow, Spark). 

Vitamin-V will address a classic cloud stack that targets the execution of entire VMs managed by OpenStack, a modern cloud setup that targets entire VMs and containers managed by Kubernetes, and a serverless cloud stack that targets the execution of lightweight VMs managed by Kubernetes with Kata Containers.

To demonstrate the three working cloud setups, they will be benchmarked against relevant AI applications (i.e., Google Net, ResBet, VGG19), Big-Data applications (TPC-DS on top of Apache Spark), and Serverless applications (FunctionBench, ServerlessBench). Vitamin-V aims to match the software performance of its x86 equivalent (x86 is the dominant ISA in cloud servers). Given the limited availability of RISC-V hardware, measures will be averaged by resorting to the CPU core mark scores to achieve a fair assessment. Overall, such development guarantees a RISC-V cloud-stack ecosystem for market take-up.

\section{Conclusions}

Cultivating a new era of open and collaborative computing, integrating RISC-V in cloud computing represents a crucial opportunity to revolutionize the industry with its unparalleled flexibility, efficiency, and scalability. Vitamin-V aims to play a crucial role in this challenging revolution. 

\printbibliography 


\end{document}